\documentclass[showpacs,twocolumn,floats,aps,prl]{revtex4}
\usepackage{times} 
\usepackage{amssymb,amsmath}
\usepackage{prelim2e}\usepackage[none,bottom]{draftcopy}
\draftcopyName{UT preprint / }{1.2} %ADAPT TEXT
\ifx\pdfoutput\undefined
\usepackage[usenames,dvips]{color}
\usepackage{graphicx}     % Include figure files
\else
\usepackage[usenames,dvipsnames]{color}
\usepackage{epstopdf}
\usepackage[pdftex]{graphicx}
\usepackage[pdftex]{hyperref}
\hypersetup{a4paper,
    pdftitle={Text on dewetting mixtures} %ADAPT TEXT 
    pdfauthor={Uwe Thiele},  %ADAPT TEXT 
pdfproducer={lateX},
pdfview=FitV,       % FitH
pdfstartview=FitB,
linkcolor=blue,     % links to same page
citecolor=blue,     % citations
pagecolor=blue,     % links to same page
urlcolor=red,      % links to URLs
breaklinks=true,    % links may be split onto 2 lines
colorlinks=true,
citebordercolor=0 0 0,  % color for \cite
filebordercolor=0 0 0,
linkbordercolor=0 0 0,
menubordercolor=0 0 0,
pagebordercolor=0 0 0,
urlbordercolor=0 0 0,
pdfhighlight=/I,
pdfborder=0 0 0,   % no box around links
backref=false,
pagebackref=false,
bookmarks=true,
bookmarksopen=true,
bookmarksnumbered=true
}
\fi
\ifx\pdfoutput\undefined
\DeclareGraphicsExtensions{.eps}
\else
\DeclareGraphicsExtensions{.jpg, .pdf, .tif, .png}
\fi
\graphicspath{{.},{./Fig/}} %$ %ADAPT DIRECTORY
\newcommand{\mylab}[1]{\label{#1}}
\renewcommand{\vec}[1]{\mathbf{#1}}
\newcommand{\vecg}[1]{\boldsymbol{#1}}

\begin{document}
%
%----------------------------------------------------------------%
\title{Gradient dynamics description for films of mixtures and
  suspensions - dewetting triggered by coupled film height and
  concentration fluctuations}
\author{Uwe Thiele}
\email{u.thiele@lboro.ac.uk}
\homepage{http://www.uwethiele.de}
\author{Desislava V. Todorova}
\author{Hender Lopez}

\affiliation{Department of Mathematical Sciences, Loughborough University,
Loughborough, Leicestershire, LE11 3TU, UK}
%\affiliation{affil}
%
\begin{abstract}
 A thermodynamically consistent gradient dynamics model for
    the evolution of thin layers of liquid mixtures, solutions and
    suspensions on solid substrates is presented which is based on a
    film height- and mean concentration-dependent free energy
    functional.  It is able to describe a large variety of structuring
    processes including coupled dewetting and decomposition
    processes. As an example, the model is employed to investigate
  the dewetting of thin films of liquid mixtures and suspensions under
  the influence of effective long-range van der Waals forces that
  depend on solute concentration. The occurring fluxes are discussed
  and it is shown that spinodal dewetting may be triggered through the
  \textit{coupling} of film height and concentration fluctuations.
  Fully nonlinear calculations provide the time evolution and
  resulting steady film height and concentration profiles.
\end{abstract}
\pacs{
68.15.+e, % Thin films: Liquid thin films
47.50.Cd, % Modeling
68.08.-p,  % Liquid-solid interfaces
47.20.Dr %Surface-tension-driven instability 
}
\maketitle
%
%
%----------------------------------------------------------------%
%
%
Understanding the behaviour of free surface layers and drops of
  simple and complex liquids becomes increasingly important because
  the drive towards further miniaturisation of fluidic systems towards
  micro- \cite{SqQu05} and eventually nano-fluidic \cite{MEv05}
  devices depends on our ability to gain control of the various
  interfacial effects on small scales. Liquid layers frequently occur
  either naturally, e.g., as tear film in the eye or industrially,
  e.g., as protection or lubrication layers.  They are also
instrumental in many wet process stages of printing,
(nano-)structuring and coating technologies where films or drops of a
liquid are applied to a surface with the aim of producing a
homogeneous or structured layer of either the liquid or a solute.  For
reviews see Refs.~\cite{ODB97,CrMa09,Bonn09,Thie10}.

Their omnipresence in natural and industrial processes provides a
strong incentive to investigate the creation, instabilities, rupture
dynamics, and short- and long-time structure formation of free surface
thin liquid films on solid substrates. These processes are well
investigated experimentally \cite{Reit92,SHJ01c} and theoretically
\cite{TVN01,Beck03} for films of simple liquids on smooth solid
substrates. Continuum models describe the evolution of the film
thickness profile $h(\vec{x},t)$ as a gradient dynamics $\partial_t h
= \vecg{\nabla}\cdot[Q(h) \vecg{\nabla} \delta F[h]/ \delta h]$ for
the free energy $F[h]=\int d\vec{x} (\gamma \xi +f(h))$
\cite{ttl13_note0} that accounts for wettability through the local
wetting energy $f(h)$ and for capillarity through the local surface
energy $\gamma \xi$ \cite{Thie10}.  Here, $\xi d\vec{x}
=(1+\frac{1}{2}|\vecg{\nabla} h|^2)d\vec{x}$ is the long-wave (or
small-gradient) approximation of the surface area
element in Monge parametrization, $\gamma$ is
the liquid-gas interface tension, the variational derivative $\delta
F[h]/\delta h=-\gamma\Delta h - \Pi(h)$ corresponds to the pressure
where $\Pi(h)=-df/dh$ is the Derjaguin or disjoining pressure
\cite{deGe85,StVe09,Isra11}, $Q(h)=h^3/3\eta$ is the mobility function
in the case of no-slip at the substrate where $\eta$ is the dynamic
viscosity (for the case of slip see, e.g., \cite{MWW05}),
$\vec{x}=(x,y)^T$, and $\vecg{\nabla}=(\partial_x,\partial_y)^T$. The
described model may be derived via a long-wave approximation from the
Navier-Stokes and continuity equations with adequate boundary
conditions at the free surface and the solid substrate
\cite{ODB97,Thie07,CrMa09}.  

The dynamics of films of simple liquids is rather well
understood. However, the situation strongly differs for films of
complex liquids as, for instance, colloidal (nano-)particle
suspensions, mixtures, polymer and surfactant solutions, polymer
blends and liquid crystals.  Practically, layers of such complex
liquids occur far more widely than films of simple liquid, but a
systematic understanding of the possible pathways of their evolution
that result from the coupled processes of dewetting, decomposition,
evaporation and adsorption has not been reached.  Free surface films
of such liquids occur, for instance, as tear films \cite{ShRu85b},
lung lining \cite{Grot11}, in the production of organic solar cells
\cite{Heie08} or semiconductor nanoparticle rings \cite{MDSY99}.
Layers of solutions and suspensions with
volatile solvent are frequently employed in intermediate stages of the
production of homogeneous or structured layers of the solute, e.g., as
a non-lithographic technique for covering large areas with regular
arrays of small-scale structures. 
Reviews of experiments, models and
  applications can be found in \cite{MaCr09} (surfactant solutions),
\cite{FAT12,HaLi12} (deposition processes from solution) and
\cite{GeKr03} (polymer blends). Although in all these systems the
interfacial effects of capillarity and wettability are still main
driving forces, they may now interact with the dynamics of inner
degrees of freedom as, e.g., the diffusive transport of solutes or
surfactants, phase separation and other phase transitions,
evaporation/condensation of solvent and concentration-dependent
wettability.

The present work provides a consistent framework for the theoretical
description of many of the observed dynamical processes in films of
liquid mixtures, solutions and suspensions. After introducing the
model, we discuss limiting cases  and
elucidate the physical meaning of the occurring fluxes. As an example
we apply the presented general framework to the case of a film of a
liquid mixture where the wettability depends on the local
concentration. This shows that dewetting may be triggered through the
\textit{coupling} of film height and concentration fluctuations.

\begin{figure}
\includegraphics[width=0.7\hsize]{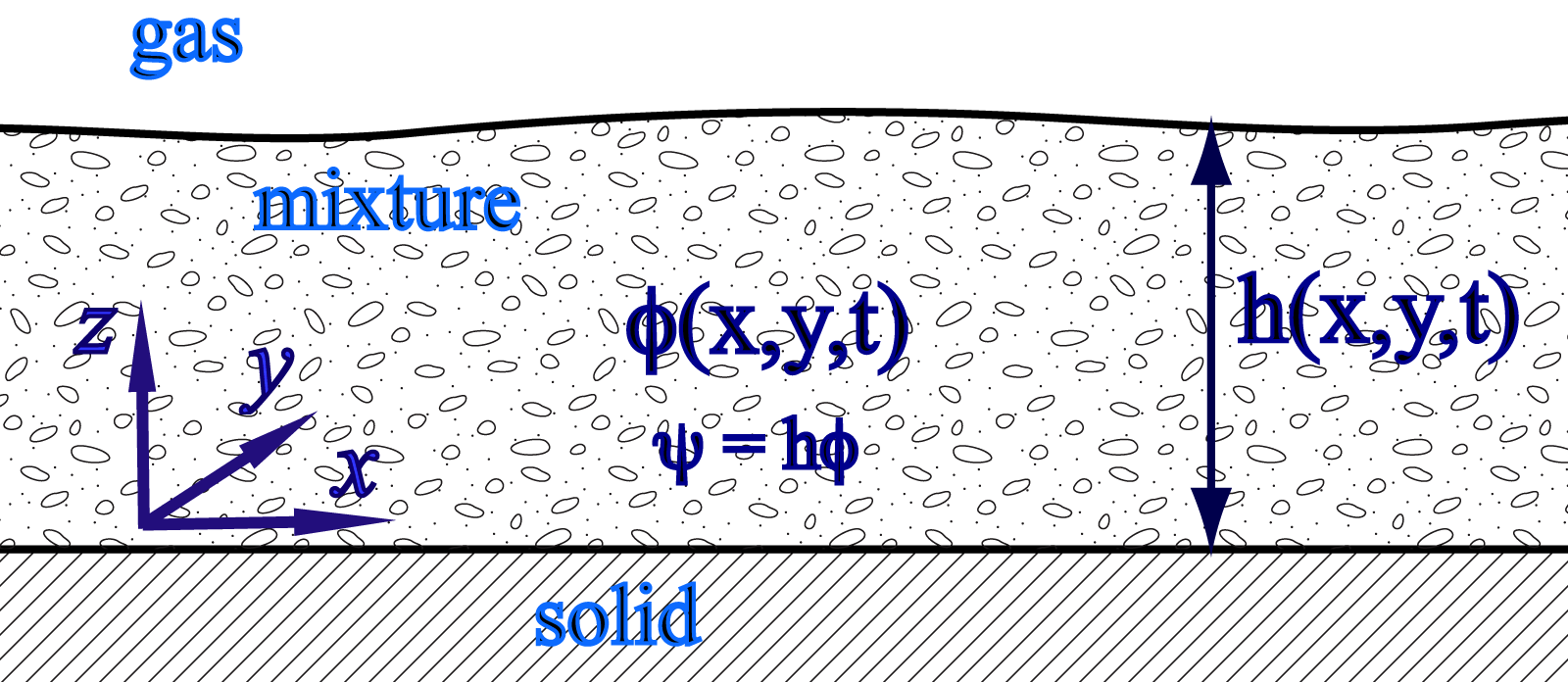}
  \caption{Sketch of the considered geometry for a film of a liquid
    mixture whose components we call solvent and solute. The relevant
    \textit{conserved} fields are the film height profile $h$ and the effective
    local solute layer thickness $\psi=h\phi$, where $\phi$ is the \textit{non-conserved}
    height-averaged solute concentration.}
\label{fig:sketch}    
\end{figure}

We consider a thin non-volatile liquid film of a mixture on a
solid substrate (see Fig.~\ref{fig:sketch}) that without additional
influx of energy relaxes towards some static equilibrium state much as
in many of the experiments reviewed in \cite{GeKr03}.  In the case
without evaporation the approach to equilibrium for this relaxational
system is described by a gradient dynamics of the underlying
  free energy functional
\begin{equation}
F[h,\psi]\,
=\,\int\left[\gamma \xi +
  f\left(h,\phi\right) + h \,g\left(\phi\right) + \Sigma\right] dA.
\mylab{eq:en3}
\end{equation}
It is an extension of the above introduced $F[h]$ that accounts for
(i) a dependence of the wetting energy on local concentration, (ii)
the bulk free energy of the mixture per substrate area $h g(\phi)$,
and (iii) the energetic cost of strong gradients in the concentration
(through $\Sigma=\frac{\sigma}{2}h |\vecg{\nabla}\phi|^2$
where $\sigma$ is the interfacial stiffness).

The gradient dynamics has to be written in terms of the \textit{conserved
  fields}, film thickness $h(\vec{x},t)$ and effective local solute
layer thickness $\psi(\vec{x},t)=h(\vec{x},t)\phi(\vec{x},t)$.  The
non-conserved field $\phi$ is the dimensionless height-averaged per
volume solute concentration. The general coupled evolution equations
for two such conserved order parameter fields in the framework of
linear nonequilibrium thermodynamics are
\begin{eqnarray}
\partial_t h \,&=&\,
\vecg{\nabla}\cdot\left[Q_{hh}\vecg{\nabla}\frac{\delta F}{\delta h}\,+\,Q_{h\psi}\vecg{\nabla}\frac{\delta F}{\delta \psi}\right],
\nonumber\\
\partial_t \psi \,&=&\,
\vecg{\nabla}\cdot\left[Q_{\psi h}\vecg{\nabla}\frac{\delta F}{\delta h}\,+\,Q_{\psi \psi}\vecg{\nabla}\frac{\delta F}{\delta \psi}\right].
\mylab{eq:solsusp-coup-grad2}
\end{eqnarray}
The mobility matrix
\begin{equation}
\mathbf{Q}\,=\,\left( 
\begin{array}{cc}  
Q_{hh} & Q_{h\psi} \\[.3ex]
Q_{\psi h} &Q_{\psi \psi}
\end{array}
\right)
\,=\,\frac{1}{3\eta}\left( 
\begin{array}{cc}  
h^3 & h^2\psi \\[.3ex]
h^2\psi \phantom{x}& h\psi^2+ 3 \eta \widetilde D \psi
\end{array}
\right)
\mylab{eq:solsusp-mob}
\end{equation}
is symmetric and positive definite corresponding to Onsager
reciprocal relations 
and the condition for positive entropy production, respectively
\cite{deMa84}. $\widetilde D $ is the molecular mobility of the solute.

To perform the variations in
Eqs.~(\ref{eq:solsusp-coup-grad2}) one has to replace $\phi$
everywhere by $\psi/h$. The extended free energy
(\ref{eq:en3}) for a film of a mixture \cite{ttl13_note0} results in convective and
diffusive fluxes (for brevity, written in terms of $h$ and $\phi$)
\begin{eqnarray}
\vec{J}_\mathrm{conv}\,&=&\frac{h^3}{3\eta}\left\{
\gamma\vecg{\nabla}\vecg{\Delta} h 
- \vecg{\nabla}\partial_h f 
+ \frac{\partial_\phi f}{h} \vecg{\nabla} \phi
\right.\mylab{eq:fluxes:conv}\\
&&\left.
-\frac{\sigma}{h}[\vecg{\nabla}\cdot(h\vecg{\nabla}\phi)]\vecg{\nabla}  \phi  
- \frac{\sigma}{2}\vecg{\nabla}|\vecg{\nabla}\phi|^2
\right\},  \nonumber \\
\vec{J}_\mathrm{diff}\,&=&\,-\widetilde D h\phi
\vecg{\nabla}\left[
\frac{\partial_\phi f}{h} + \partial_\phi g - \frac{\sigma}{h}\vecg{\nabla}\cdot(h\vecg{\nabla}\phi)
\right],
\mylab{eq:fluxes:diff}
\end{eqnarray}
respectively. Employing the fluxes we
bring the gradient dynamics equations (\ref{eq:solsusp-coup-grad2})
into the form
\begin{eqnarray}
\partial_th &=& -\vecg{\nabla}\cdot\vec{J}_\mathrm{conv},
\mylab{e:fluxform1}\\
\partial_t(\phi h) &=& -\vecg{\nabla}\cdot(\phi\vec{J}_\mathrm{conv}
+ \vec{J}_\mathrm{diff}),
\mylab{e:fluxform2}
\end{eqnarray}
which is common in the hydrodynamic literature \cite{ODB97,MaCr09,FAT11}.

Before discussing important limiting cases, we elucidate the physical
meaning of the individual flux contributions. In the convective flux
[Eq.~(\ref{eq:fluxes:conv})] the first term is due to Laplace pressure
gradients \cite{ODB97}; the second term is the Derjaguin pressure
contribution due to wettability; and the final two terms represent the
Korteweg flux, i.e., a bulk concentration-gradient driven flux
(cf.~\cite{AMW98} for a discussion of the related bulk model-H). The
third term is a flux driven by concentration-gradients within the bulk
of the film but only if the film is sufficiently thin such that its
two interfaces 'feel' each other. This novel flux is a direct
consequence of the concentration dependence of the wetting energy
$f(h,\phi)$ and has a similar magnitude as the Derjaguin pressure
contribution \cite{TTL13_note1}.

The first term of the diffusive flux [Eq.~(\ref{eq:fluxes:diff})] is
also uncommon in the literature although it is a natural consequence
of the gradient dynamics form (\ref{eq:solsusp-coup-grad2}).  It
represents the influence of the concentration-dependent wettability on
diffusion. The second term is the flux due to gradients of the
chemical potential $\mu=\partial_\phi g$ in the bulk of the film while
the final term is a Korteweg contribution to diffusion that counters
steep concentration gradients, e.g., for decomposing solvent-solute
films.

The general evolution equations
[(\ref{e:fluxform1},\ref{e:fluxform2}) with
(\ref{eq:fluxes:conv},\ref{eq:fluxes:diff})] recover several known
models as limiting cases (this is used to determine $\mathbf{Q}$).
Most importantly: (i) For a constant film height $h$, without
wettability contribution ($f=0$) and appropriately defined $g$,
Eq.~(\ref{e:fluxform2}) becomes the Cahn-Hilliard equation that
describes, e.g., the spinodal decomposition of a binary mixture
\cite{Cahn65}; (ii) As in (i) but with $\sigma=0$ and a purely
entropic (ideal gas-like)
\begin{equation}
g=g_\mathrm{id}(\phi)=\frac{k_B T}{a^3}\, \phi (\log \phi - 1),
\mylab{eq:en4}
\end{equation}
where $a$ is a molecular length scale related to the solute, one
recovers the standard diffusion equation with diffusion constant
$\widetilde D k_BT/a^3$ (see, e.g., section IV of \cite{TAP12});
(iii) For $f=f(h)$, $\sigma=0$ and $g=g_\mathrm{id}$ one recovers the
conserved part of long-wave equations used, e.g., to study dewetting
of and solute deposition from solutions and suspensions
\cite{WCM03,FAT11,ttl13_note3}; 
(iv) Again without wettability, but with Korteweg fluxes
($\sigma\neq0$), and employing the double-well potential
$g\sim(\phi^2-1)^2$ for the solvent-solute interaction one obtains the
thin film limit of model-H \cite{AMW98} as derived recently via a
long-wave asymptotic expansion \cite{NaTh10,TTL13_note2}.

Next we present as an example the practically relevant case
of a solute-dependent wettability, i.e., $f=f(h,\phi)$. For clarity we
only include entropic bulk terms for the solute-solvent interaction,
i.e.\ $g=g_\mathrm{id}$ [Eq.~(\ref{eq:en4})] and $\partial_\phi g=(k_B
T/a^3) \log \phi$, implying absolute stability against bulk
solute-solvent decomposition.  For the wetting energy we use the
combination of long-range van der Waals interactions and an always
stabilising ($B > 0$) short-range contribution \cite{Isra11,Pism02}:
\begin{equation}
f(h,\phi) = -\frac{A(\phi)}{2h^2} + \frac{B}{5h^5}.
\end{equation}

\begin{figure}
\textbf{(a)}\hspace{-0.04\hsize}\includegraphics[width=0.5\hsize]{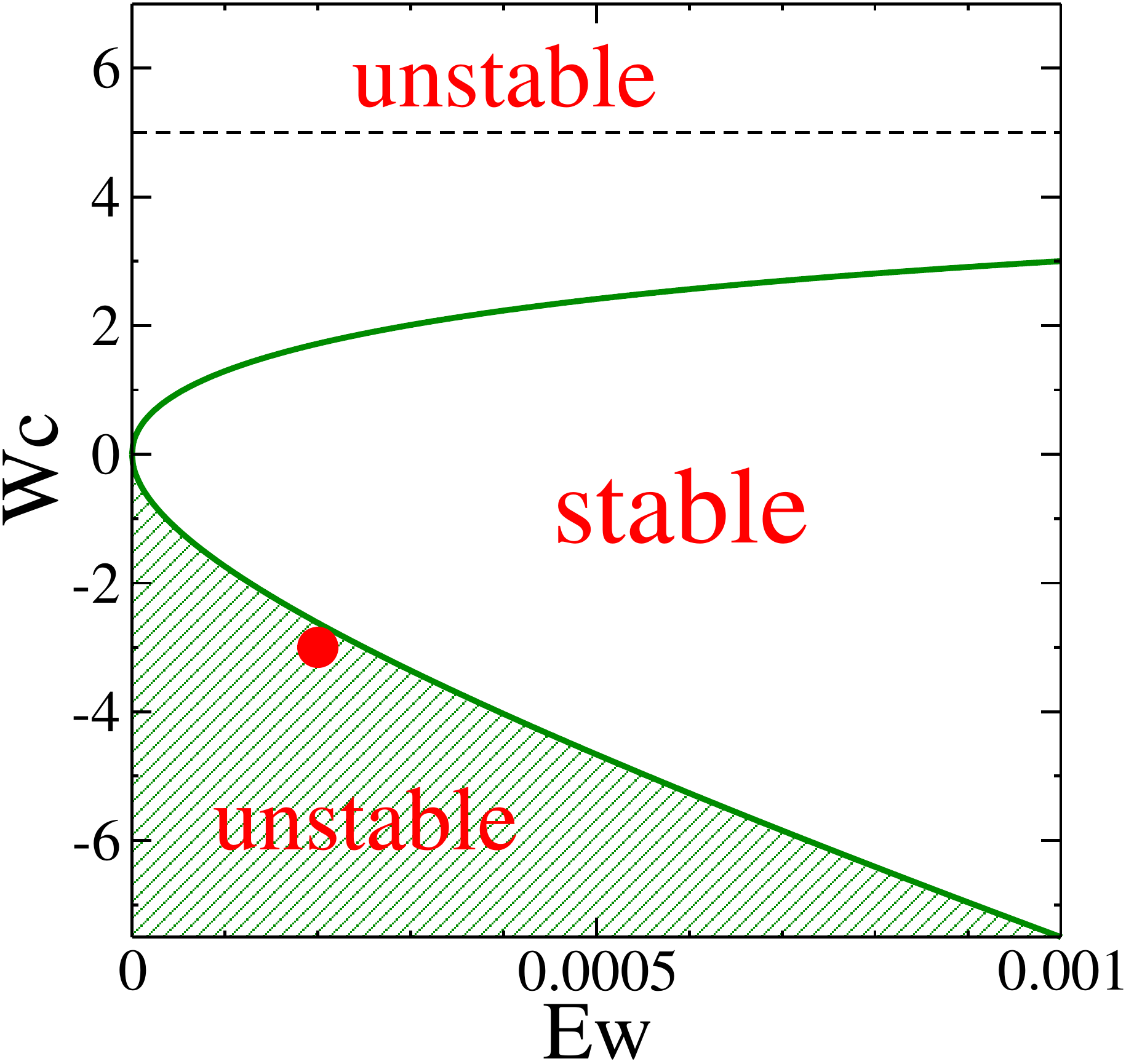}
\hspace{0.05\hsize}
\begin{minipage}[b]{0.4\hsize}
\includegraphics[width=0.95\hsize]{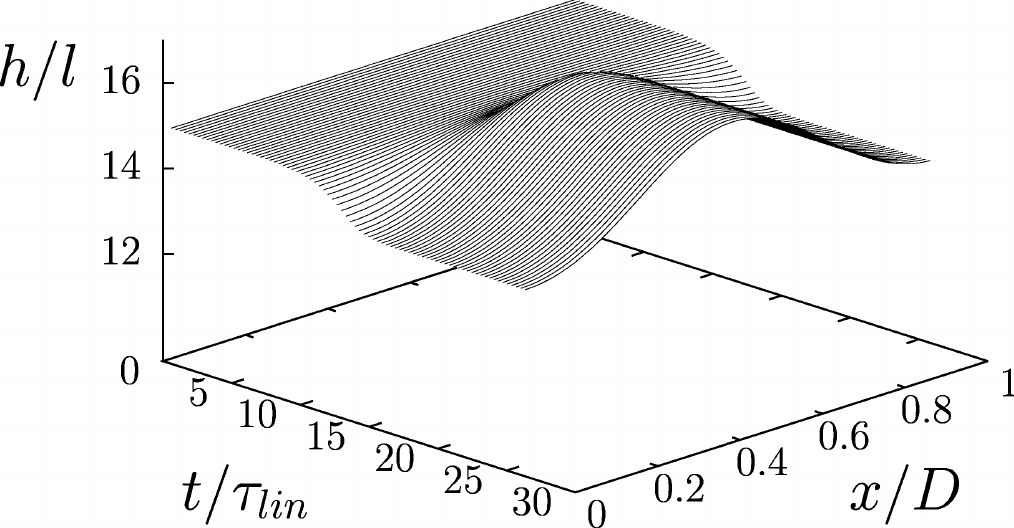}\textbf{(b)}\\[3ex]
\includegraphics[width=0.95\hsize]{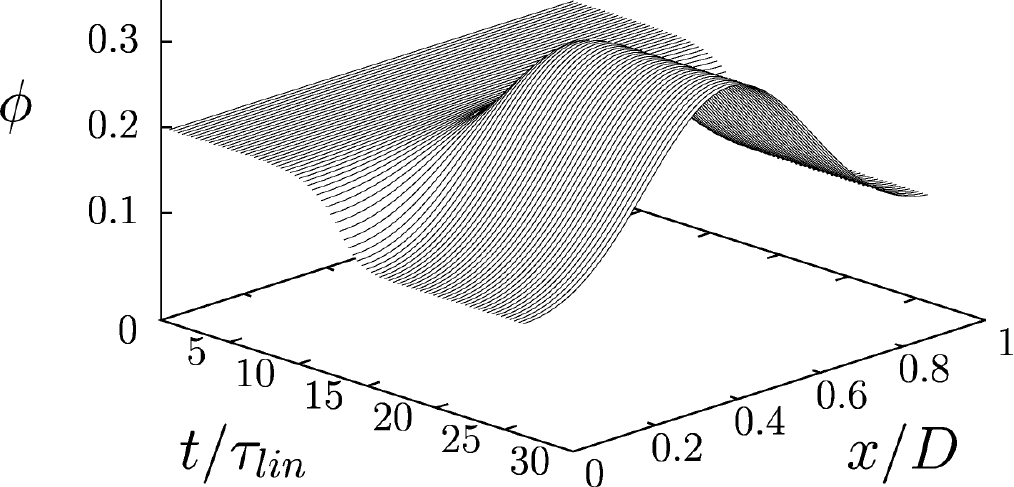}\textbf{(c)}\\[1ex]
\end{minipage}
\caption{(Color online) Shown is in (a) the linear stability of flat
  homogeneous films (of thickness $h_0/l=15$ and concentration
  $\phi_0=0.2$) with respect to coupled fluctuations in film height
  and concentration in the plane spanned by the ratio of entropic and
  wetting influences $\mathrm{Ew}$ and the strength of the
  concentration-dependence of wettability $\mathrm{Wc}$. For parameter
  values $\mathrm{Ew}=0.002$ and $\mathrm{Wc}=-3$ marked by the red
  circle in (a), panels (b) and (c) show for the case of one spatial
  dimension (dimensionless domain size $D/L=1500$) space-time plots of
  the fully nonlinear coupled short-time evolution of the height and
  concentration profile, respectively. Time is given in units of the
  typical time $\tau_\mathrm{lin}$ of the fastest linear instability
  mode. In all calculations, the nondimensional interfacial stiffness
  is fixed to $\sigma / (l \gamma)=0.1$ while the nondimensional
  surface tension is set to one through the choice of the lateral
  lengthscale $L$ \cite{TTL13_note_scales}.  }
\label{fig:linstab}    
\end{figure}

Note that we combine a concentration-dependent Hamaker 'constant'
$A(\phi)$ and a constant $B$. One may as well introduce a
concentration-dependent short-range contribution or use a different
form for the short-range contribution \cite{TTL13_djpress}, however, these choices do
not affect the main results.  The Derjaguin pressure is
$\Pi(h,\phi)=-\partial_h f$ while $\partial_\phi f$ could be called a
Derjaguin chemical potential. $A(\phi)$ is determined employing
homogenization techniques. For many experimentally employed mixtures
as e.g., PMMA/PS, toluene/acetone or PS/toluene on Si or SiO a linear
dependence is an excellent approximation over the entire concentration
range \cite{Todo13}.
Selecting the case where the pure solvent is wetting $A_0\equiv
A(\phi=0) < 0$, we write $A(\phi)=|A_0|(-1+\mathrm{Wc}\, \phi)$ where
the nondimensional number $\mathrm{Wc}$ quantifies the strength of the
concentration-dependence of wettability.  Experimentally,
$\mathrm{Wc}$ may be changed by choosing a different solute. For the
materials we are interested in, $|A_0|$ varies in the range
$[10^{-22}, 10^{-19}]$ Nm and $\mathrm{Wc}$ lies in the range $[-10,
15]$. For example, a mixture of polystyrene (PS) and poly(methyl
methacrylate) (PMMA) on a silicon (Si) substrate (used, e.g., in
\cite{HeJo05}) yields $A_0 = -2.22\times10^{-19}$ and $\mathrm{Wc} =
-0.11$ and for a solution of PS in toluene on silicon oxide (SiO) one
obtains $A_0 = 1.74\times10^{-21}$ and $\mathrm{Wc} = 7.1$, while a
mixture of toluene and acetone on SiO gives $\mathrm{Wc} = -9.0$.

Note that for $A_0<0$ and $\mathrm{Wc}<0$, both, a film of pure
solvent and a film of pure solute, are absolutely stable. With
$g=g_\mathrm{id}$ the bulk solute-solvent mixture is stable as well. A
film of mixture might then be expected to be stable for all
$\mathrm{Wc} < 0$ and to become unstable for $\mathrm{Wc} > 0$ when
$\mathrm{Wc}\, \phi>1$ because then $A(\phi)>0$. This expectation,
however, assumes that the mixture in the film remains homogeneous,
i.e., that concentration fluctuations are always damped.
However, a linear stability analysis of flat homogeneous films with
respect to fluctuations $\delta h$ and $\delta\psi$ shows that the
fluctuations in film height and concentration couple, rendering the
system more unstable. Fig.~\ref{fig:linstab}(a) shows that even for
$\mathrm{Wc} < 0$ where all decoupled subsystems are stable, the film
of a mixture can be linearly unstable in an extended experimentally
accessible range of the parameter space.

Here the dimensionless number $\mathrm{Ew}=k_bTl^3/|A_0|a^3$ is the
ratio of entropic and wetting influences \cite{TTL13_note_scales}. In
other words a film of stable solvent can be destabilized by a stable
solute if the diffusion of the solute is sufficiently weak, i.e., if
$\mathrm{Ew}$ is sufficiently small. For common mixtures,
  solutions and nanoparticle suspensions $\mathrm{Ew}$ can range from
  $O(10^{-4})$ to $O(10^{4})$. The estimate is based on a typical
  precursor thickness of $l\sim1$~nm \cite{PODC12} and a solute length
  scale $a$ between $0.1$~nm and $10$~nm (this corresponds e.g.~to the
  size of molecules or (nano-)particles diffusing in the film)
  \cite{Todo13}.
Also for $\mathrm{Wc} > 0$ the film becomes unstable at smaller
$\mathrm{Wc}$ than expected under the assumption that the mixture
stays homogeneous (dashed line in Fig.~\ref{fig:linstab}(a)). Because
$h$ and $\psi$ are both conserved, the instability is of long
wavelength, i.e., at onset (at critical $\mathrm{Wc}$ or
$\mathrm{Ew}$) it has zero wavenumber (cf.~\cite{Thie10}). Therefore,
for finite domains the stability borders in Fig.~\ref{fig:linstab}(a)
are slightly shifted.

\begin{figure}
\includegraphics[width=0.45\hsize]{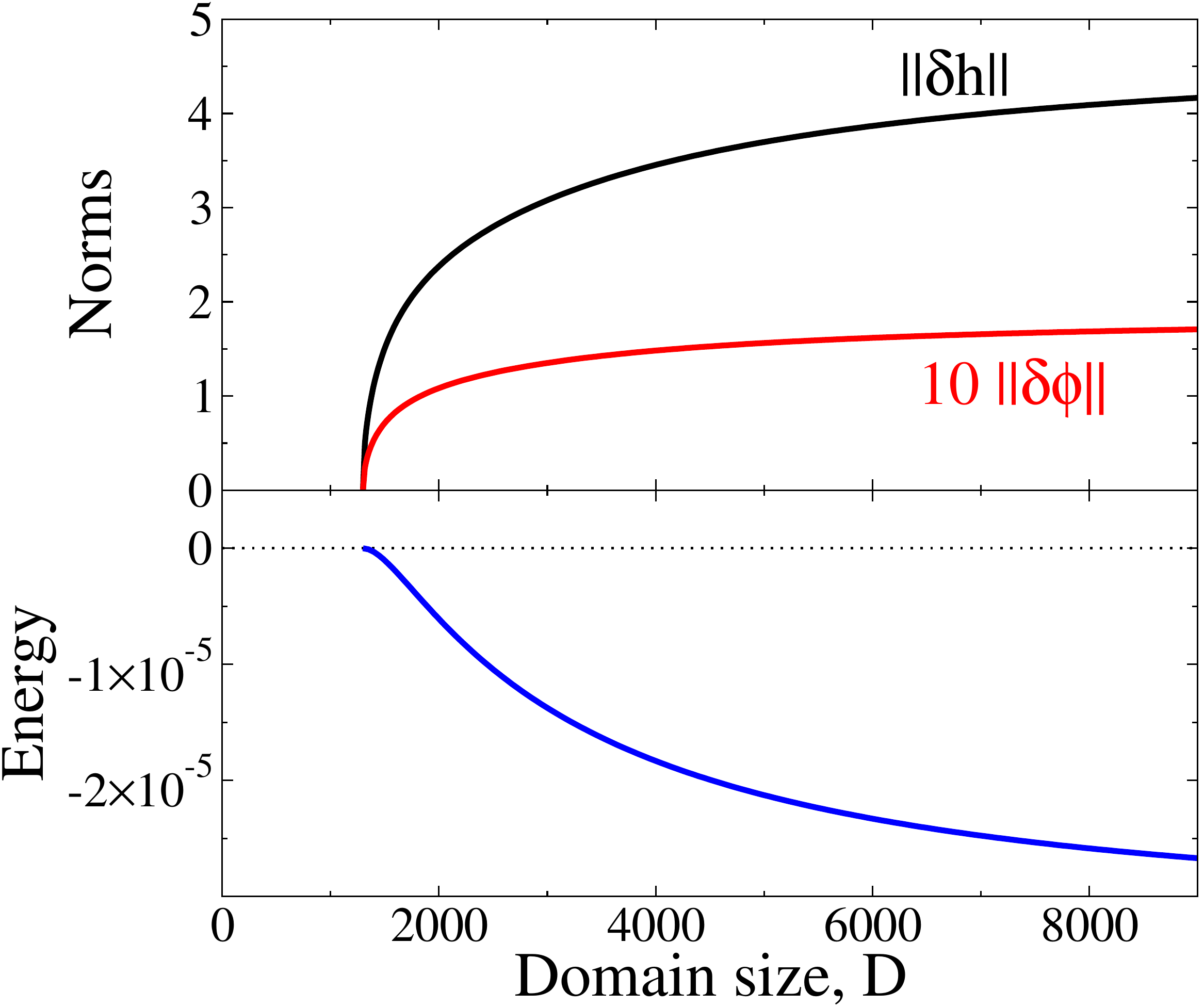}
\includegraphics[width=0.45\hsize]{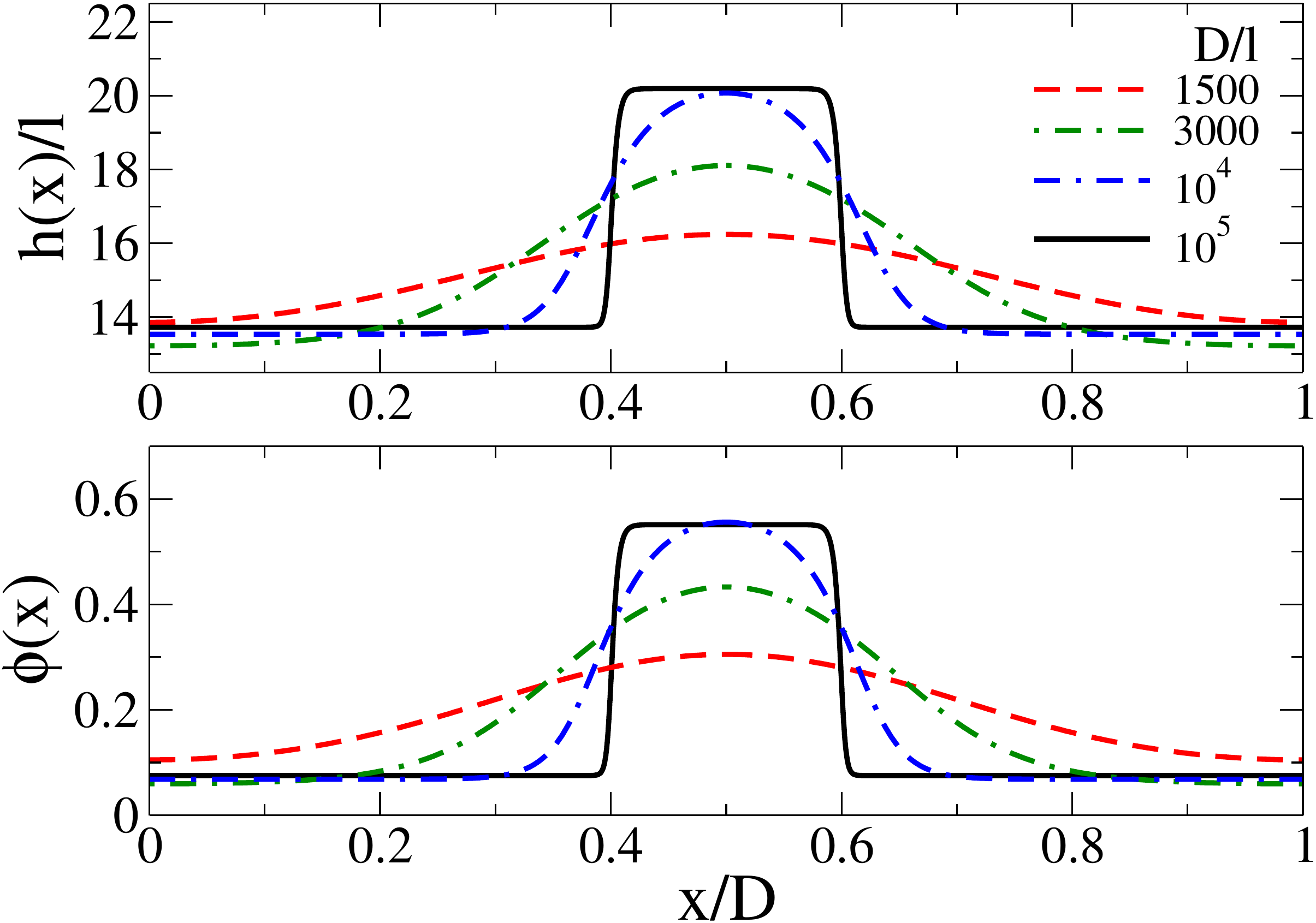}
\caption{(Color online) (a) Norms and (b) energy per length of the
  family of steady drops in dependence of the domain
  size $D/L$ at $\mathrm{Ew}=0.002$, $\mathrm{Wc}=-3$ corresponding to
  the red square in Fig.~\ref{fig:linstab}(a). (c) Thickness (top) and
  concentration (bottom) profiles at various domain sizes as given in
  the legend.}
\label{fig:prof}    
\end{figure}

Starting from a homogeneous flat film, we illustrate in
Fig.~\ref{fig:linstab}(b,c) the resulting spontaneous structure
formation \cite{TTL13_num}. During the shown linear and nonlinear
stages of the short-time evolution, the steady state shown in
Fig.~\ref{fig:prof}(b) is approached \cite{TTL13_num}. In a large
domain many such small droplets will undergo a long-time coarsening
process (not shown) to reach pancake-like drops as shown in
Fig.~\ref{fig:prof}(b) for $D/L=10^5$. Inspecting the $h$ and $\phi$
profiles and the energy in Fig.~\ref{fig:prof} the physical mechanism
that drives the structuring becomes clear: Although the film can not
reduce its energy by modulating its thickness profile at homogeneous
concentration, it is still able to do so by simulaneously modulating
its thickness and concentration profiles. In the present example the
solute is enriched [depleted] in the thicker [thinner] part of the
profile. The characteristics of the coexisting flat parts visible in
Fig.~\ref{fig:prof}(b) for $D/L=10^5$ may also be obtained through an
analysis of the binodals of the system, i.e., of the film height and
concentration values at coexistence \cite{Todo13}. Note that the
structuring results in extended flat regions of different heights that
are still much larger than the vertical lengthscale, i.e., all regions
may still accommodate a diffusing solute with $a>l$. Furthermore, one
may include steric effects due to the solute size into the free
energy.

The presented example illustrates that the above introduced
thermodynamically consistent long-wave model allows one to predict a
novel interface instability for thin films (below about 100nm) of
liquid mixtures and suspensions under the influence of long-range van
der Waals forces that are concentration dependent. The resulting
\textit{coupling} of film height and concentration fluctuations always
renders such films more unstable than the decoupled subsystems. The
chosen numerical example shows that the destabilization can even occur
if all decoupled subsystem are unconditionally stable. However, the
presented gradient dynamics formulation has further far reaching
implications for the description of thin film of complex fluids: For
so-called nanofluids (nano-particle suspensions) often a structural
Derjaguin pressure \cite{DeCh74} is included into the hydrodynamic
description \cite{WaNi03,CMS09}. However, Eqs.~(\ref{eq:fluxes:diff})
shows that this is incomplete. Instead, a structural wetting energy
has to be introduced what results, in consequence, in additional
contributions to the convective, diffusive (and evaporative) flux. An
accounting for attractive solvent-solute interactions (beyond the
entropic term considered in the example) allows one to investigate how
the various decomposition and dewetting instability modes couple,
resulting in a number of different instability types and evolution
pathways somewhat similar to the ones described for two-layer films of
immiscible liquids \cite{PBMT05}.

In summary, we have presented a general gradient dynamics model and a
particular underlying free energy \cite{ttl13_note0} which is able to
describe a wide range of dynamical processes in thin films of liquid
mixtures, solutions and suspensions on solid substrates including the
dynamics of coupled dewetting and decomposition. We have argued that
on the one hand the model recovers known limiting cases including the
long-wave limit of model-H. On the other hand we have discussed the
physical meaning of important contributions that are missing in the
hydrodynamic literature, and have shown that they are needed for a
thermodynamically consistent description of, e.g., evolution pathways
controlled by concentration-dependent wettability. As an example, we
have investigated the dewetting of thin films of liquid mixtures and
suspensions under the influence of long-range van der Waals forces
that are concentration dependent.

The presented gradient dynamics form will allow for systematic future
developments. Most importantly, the here presented model for a film of
a mixture \textit{without} enrichment or depletion boundary layers at
the interfaces may be combined with models for films with an insoluble
surfactant \cite{KGFC10,TAP12} to also describe systems where
enrichment or depletion layers form at the interfaces, including
instabilities and structuring processes as observed in \cite{Thom10}.

\acknowledgments

This work was supported by the European Union under grant
PITN-GA-2008-214919 (MULTIFLOW).

%%%%%%%%%%%%%%%%%%%%%%%%%%%%%%%%%%%%%%%%%%%%%%%%%%%%%%%%%%%%%%%%%%%%%%%%%%%%%%%

%

\begin{thebibliography}{48}
\bibitem[{ttl()}]{ttl13_note0}
\bibinfo{note}{Note, that $F[h]$ is sometimes called an ''effective interface
  Hamiltonian'' (see~\cite{Bonn09}), but could equally be called a
  ''Lagrangian'' (based on an analogue to point mechanics).}
\bibitem[{ttl()}]{ttl13_note3}
\bibinfo{note}{Note, that adding a non-conserved $\sim\delta F/\delta h$ term
  to the first equation of (\ref{eq:solsusp-coup-grad2}) accounts for the
  evaporation of a volatile solvent including the osmotic pressure dependence
  missed in \cite{WCM03,FAT11}.}
\bibitem[{TTL({\natexlab{a}})}]{TTL13_note2}
\bibinfo{note}{A small discrepancy in the mobilities arises because the free
  energy and the mobility, on which the Cahn-Hilliard Equation is based, are
  approximated independently of each other.}
\bibitem[{TTL({\natexlab{b}})}]{TTL13_note_scales}
\bibinfo{note}{For the analysis,
  Eqs.~(\ref{eq:solsusp-coup-grad2})-(\ref{eq:en3}) are nondimensionalised
  employing scales $l=(B/|A_0|)^{1/3}$, $L=l^2(\gamma/|A_0|)^{1/2}$, and
  $\tau=l\eta/\kappa$ for vertical and horizontal length scales and time,
  respectively, where the energy scale is $\kappa=|A_0|/l^2$. In the case of a
  partially wetting solvent, $l$ corresponds to the precursor film height.}
\bibitem[{TTL({\natexlab{c}})}]{TTL13_djpress}
\bibinfo{note}{Note that other combinations of long-range destabilizing
  and short-range stabilizing terms in the wetting energy may be used to
  describe partially wetting liquids on a solid substrate (see, e.g.,
  Refs.~\cite{SHJ01c,Thie07,Isra11}). Here, we employ a combination that arises
  as sharp interface limit of an underlying density functional theory based on
  Lenard-Jones molecular interactions with an $1/r^6$ attraction between
  molecules and hard-core repulsion \cite{Pism02}. Note, however, that the
  system behaviour is very similar if wetting energies of a similar form are
  chosen that result in identical precursor film heights and equilibrium
  contact angles.}
\bibitem[{TTL({\natexlab{d}})}]{TTL13_num}
\bibinfo{note}{For the time simulations we use a second order central finite
  difference scheme in space and a variable-order and variable-step backward
  differentiation algorithm in time. Periodic boundary conditions are used and
  the simulation is initiated with flat $h$ and $\psi$ profiles perturbed by a
  small uniform noise. Steady profiles and bifurcation diagrams are numerically
  obtained employing pseudo-arclength continuation.}
\end{thebibliography}
\end{document}